\documentclass{aa}  

\usepackage{revsymb}
\usepackage{graphicx}
\usepackage{amsmath,amssymb}
\usepackage{xcolor}
\usepackage{natbib}
\usepackage{textcomp}

\bibpunct{(}{)}{;}{a}{}{,}             
\catcode`\@=11
\def\gsim{\ifmmode{\mathrel{\mathpalette\@versim>}}
    \else{$\mathrel{\mathpalette\@versim>}$}\fi}
\def\lsim{\ifmmode{\mathrel{\mathpalette\@versim<}}
    \else{$\mathrel{\mathpalette\@versim<}$}\fi}
\def\@versim#1#2{\lower 2.9truept \vbox{\baselineskip 0pt \lineskip
    0.5truept \ialign{$\m@th#1\hfil##\hfil$\crcr#2\crcr\sim\crcr}}}
\catcode`\@=12

\begin{document} 

 \title{Proper motions in the VVV Survey: Results for more than 15 million 
stars across NGC~6544
   \thanks{Based on observations taken with ESO telescopes at 
   Paranal Observatory under programme IDs 179.B-2002}}

\author{
R. Contreras Ramos\inst{1,2} 
 \and
M. Zoccali\inst{1,2}
\and
F. Rojas\inst{2}
\and
A. Rojas-Arriagada\inst{1,2}
\and
M. G\'arate\inst{2,5}
\and           
P. Huijse\inst{1,6}
\and           
F. Gran\inst{2}
\and
M. Soto\inst{7}
\and
A.A.R. Valcarce\inst{2}
\and
P. A. Est\'evez\inst{1,6}
\and
D. Minniti\inst{1,3,4}
}

\institute{
Millennium Institute of Astrophysics, Av. Vicu\~na Mackenna 4860, 782-0436 Macul, 
Santiago, Chile
\email{rcontrer@astro.puc.cl}
\and
Instituto de Astrof\'isica, Pontificia Universidad  Cat\'olica de Chile, Av. 
Vicu\~na Mackenna 4860, 782-0436 Macul, Santiago, Chile
\and
Departamento de Ciencias F\'isicas, Universidad Andr\'es Bello, Campus La Casona, 
Fern\'andez Concha 700, Santiago, Chile
\and
Vatican Observatory, Vatican City State V-00120, Italy
\and
Millennium Nucleus ``Protoplanetary Disks'', Santiago, Chile
\and
Departamento de Ingenier\'ia El\'ectrica, Universidad de Chile, Av. Tupper 2007, 
Santiago, Chile
\and
Universidad de Atacama, Departamento de F\'isica, Copayapu 485, Copiap\'o, Chile
}


  \abstract
{In  the last  six  years, the  VISTA
Variable   in  the   V\'\i  a   L\'actea  (VVV)   survey   mapped  562
sq. deg. across the bulge and  southern disk of the Galaxy. However, a
detailed  study of these  regions, which  includes $\sim$36 globular
clusters (GCs) and  thousands of open clusters is by  no means an easy
challenge.  High differential reddening  and severe crowding along the
line  of  sight makes  highly  hamper  to  reliably distinguish  stars
belonging to different populations and/or systems.}
{The  aim of this  study is  to separate
stars  that  likely  belong  to  the Galactic  GC  NGC~6544  from  its
surrounding field by means of proper motion (PM) techniques.}
{This  work  was based  upon  a  new
astrometric reduction method optimized for images of the VVV survey.}
{PSF-fitting photometry  over the six
years baseline of the survey allowed  us to obtain a mean precision of
$\sim$0.51  ~mas~yr$^{-1}$,  in each  PM  coordinate,  for stars  with
$K_{s}<15$~mag.  In the area studied here, cluster stars separate very
well from  field stars, down to  the main sequence  turnoff and below,
allowing us to derive for the  first time the absolute PM of NGC~6544.
Isochrone fitting  on the  clean and differential  reddening corrected
cluster color  magnitude diagram yields  an age of  $\sim$11$-$13 Gyr,
and metallicity [Fe/H] = $-1.5$ dex,  in agreement with previous studies
restricted to  the cluster core.  We  were able to  derive the cluster
orbit assuming an  axisymmetric model of the Galaxy  and conclude that
NGC~6544  is  likely a  halo  GC.  We  have  not  detected tidal  tail
signatures associated  to the cluster, but a  remarkable elongation in
the galactic center direction  has been found.  The precision achieved
in the  PM determination also allows  us to separate  bulge stars from
foreground disk stars, enabling the kinematical selection of bona fide
bulge stars across the whole survey area.}
{Kinematical techniques are a fundamental step toward disentangling
different  stellar populations that  overlap in  a studied  field. Our
results  show that VVV  data is  perfectly suitable  for this  kind of
analysis.}

   \keywords{ techniques: proper motions -- 
              Galaxy: kinematics and dynamics --
              Galaxy: bulge --
              Galaxy: disk --
              globular clusters:~individual:~NGC~6544
              }

\titlerunning{VVV PMs across the GC NGC~6544}
\authorrunning{Contreras Ramos et al.}

   \maketitle

\section{Introduction}

VISTA  Variables  in  the  V\'ia   L\'actea  (VVV)  is  a  public  ESO
near-infrared (NIR) survey aimed to obtain the most accurate 3D map of
the  Milky  Way  (MW)  bulge  and inner  disk,  using  variable  stars
\citep{minniti10}.

During  its   six  years  of   operation,  the  VVV   survey  observed
approximately  250 deg$^{2}$  across the  inner  disk ($294.7^{\circ}$
$\leq  l  \leq$  $350.0^{\circ}$;   $-2.25^{\circ}$  $\leq  b  \leq  $
$+2.25^{\circ}$)   and  315  deg$^{2}$   in  the   bulge  of   the  MW
($-10^{\circ}$ $\leq l \leq $ $+10.4^{\circ}$; $-10.3^{\circ}$ $\leq b
\leq $  $+5.1^{\circ}$).  Thanks  to the use  of multicolor  NIR bands
($ZYJHK_{s}$),  VVV data  can penetrate  gas  and dust  in the  plane,
exploring  yet unknown  regions  of the  MW.   At the  same time,  the
$\sim$80 epochs in $K_{s}$ provide the opportunity to search for signs
of  variability in  about  a billion  sources  \citep{saito12} in  the
region where most of the stars in the MW are confined.

The study  of the  inner regions  of the MW  is difficult  for several
reasons.  The  high and  spatially variable dust  extinction mentioned
above can be minimized using NIR bands. Crowding is also an issue that
requires relatively good  seeing ($\sim$1 arcsec for  VVV data) and a
small  pixel scale (0.34  arcsec in  VIRCAM) and  in our  case becomes
critical only for stars in  the lower main sequence.  Another problem,
however,  still  affects the  interpretation  of  the photometry:  the
presence of more than one population  along the line of sight. A large
number of foreground disk stars,  mostly in the main sequence but also
in the  red giant and core  helium burning phases, are  seen along any
line  of sight  toward  the bulge.   In  addition, the  study of  star
clusters toward the bulge is  complicated by the presence of both disk
and  bulge stars,  not easy  to tell  apart from  the  color magnitude
diagram (CMD) alone.

An efficient  way to disentangle different populations  in any stellar
field  is  by means  of  proper motion  (PM)  analysis,  a well  known
technique that helps to  isolate kinematically the contribution of the
different  stellar   components  across  the  area   under  study.  PM
measurements  are one  of  the  most promising  methods  to study  the
Galactic  bulge. As  a matter  of fact,  unlike external  galaxies the
stars in  the MW bulge  are close enough  that is possible  to resolve
them  and study  the  stellar populations  and  stellar kinematics  in
detail.  However, our location inside the disk has limited the optical
observations  mainly  to  carefully  selected  low-extinction  windows
toward  the  center, or  to  relatively  shallow  (optical) images  in
regions with poorer visibility  conditions. The first PM investigation
of the Galactic bulge was  made by \cite{spaenhauer92} and since then,
several   additional  studies   have   been  followed   \citep[][among
  others]{mendez96,sumi04,vieira07}.  Many  studies have exploited the
exquisite   precision  astrometry   provided  by   space-based  images
\citep{kuijken02,clarkson08,soto14},  however limiting  their analysis
to tiny regions in the sky.

The current  instrumentation has improved  to the point that  today is
possible  to  apply successfully  PM  techniques  to  data taken  from
ground-based  telescopes  with  relatively  short time  coverages.   A
decade  ago,  \cite{anderson06}  adapted  the  methods  developed  for
high-precision  astrometry and photometry  with the  {\it{Hubble Space
    Telescope}}  {\it(HST)}  to the  case  of wide-field  ground-based
images and described in detail the steps that are necessary to achieve
good  astrometry  with  wide-field  detectors.  To  test  the  method,
\cite{anderson06} computed  PMs for the two  closest Galactic globular
clusters  (GCs), showing  that ground-based  observations with  a time
span of a few years allowed a successful separation of cluster members
from Galactic field stars.  In the following years, the same technique
was         applied         to         more        distant         GCs
\citep{bellini09,zloczewski11,sariya12,sariya15,sariya17}   and   open
clusters    \citep{yadav08,montalto09,yadav13,bellini10}.    Recently,
\cite{libralato15} followed  the same  prescription to compute  PMs in
the field of the GCs M~22 using four years of VVV data.

The  large coverage  in space  and time  of the  VVV survey  opens the
opportunity of  using PM  techniques to study  an area across  the sky
that includes  the whole  Galactic bulge, part  of the  southern disk,
$\sim$36 globular  clusters and thousands of open  clusters.  The VVV
observations are now complete, providing  a time baseline of more than
five years for PM studies in addition to stellar variability.  To this
aim, we  have developed an automated  code to derive PMs  in VVV data,
following   the   approach    described   in   \cite{anderson06}   and
\cite{bellini14}.  As a  test of its precision, we  applied our method
to the field including the Galactic GC NGC~6544.

\begin{figure}
\centering
    {\includegraphics[angle=0,width=0.95\columnwidth]{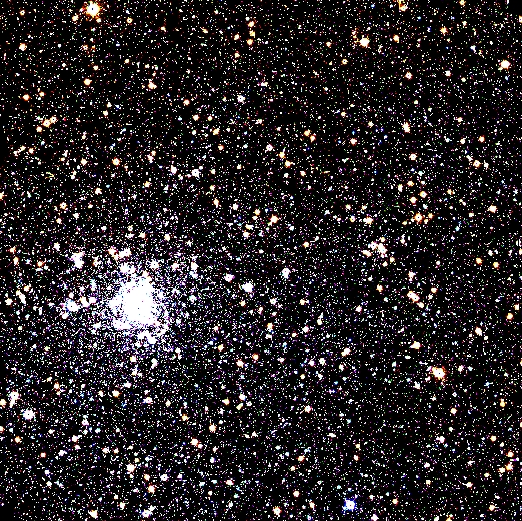}}
    \caption{Color image of chip $\#15$, in pawprint $\#2$ of
             VVV field b309. The field of view of this image is 
             $12^{\prime} \times  12^{\prime}$;  north Galactic pole is up, 
             positive galactic longitudes are left.}
    \label{FoV}
\end{figure}

NGC~6544 (see Fig.~\ref{FoV}) is a relatively metal poor \citep[Fe/H =
  $-1.4$  dex;][]{harris96}  GC located  at  a  projected distance  of
$\sim$6    kpc   from   the    Galactic   center,    specifically   at
$(l,b)=(5.84,-2.2)$.   Its position  in the  sky against  the Galactic
bulge  and disk  implies  that  a large  contribution  of field  stars
contaminate its CMD, making the identification of cluster members from
photometry alone highly uncertain.  This explains why this cluster has
been  the target  of  only  a few  photometric  and spectral  studies.
Thanks  to  the cluster  proximity  \citep[d  = 2.5  kpc,][]{cohen14},
cluster giants  are bright and easy  to separate from  field stars, in
the CMD. Below  the main sequence turnoff, however,  the only reliable
way to disentangle the different  populations is to use PMs.  To date,
the only  attempt at deriving  PMs was carried out  by \cite{cohen14}.
They computed relative  PMs between cluster and bulge  stars, in order
to select bona-fide cluster  members in the ($40^{\prime\prime} \times
40^{\prime\prime}$) area covered by  the PC/WFPC2 camera on board {\it
  HST}.  Their  purpose was to obtain accurate  physical parameters of
the cluster inner region rather than to derive precise PMs over a wide
area.   \cite{cohen14} suggest  that  the cluster  is possibly  losing
stars due to the gravitational  interaction with the MW.  This makes a
wide  area PM  study particularly  interesting, in  order to  look for
possible cluster extended tidal tails.

\begin{figure*}
\centering
\includegraphics[scale=0.21]{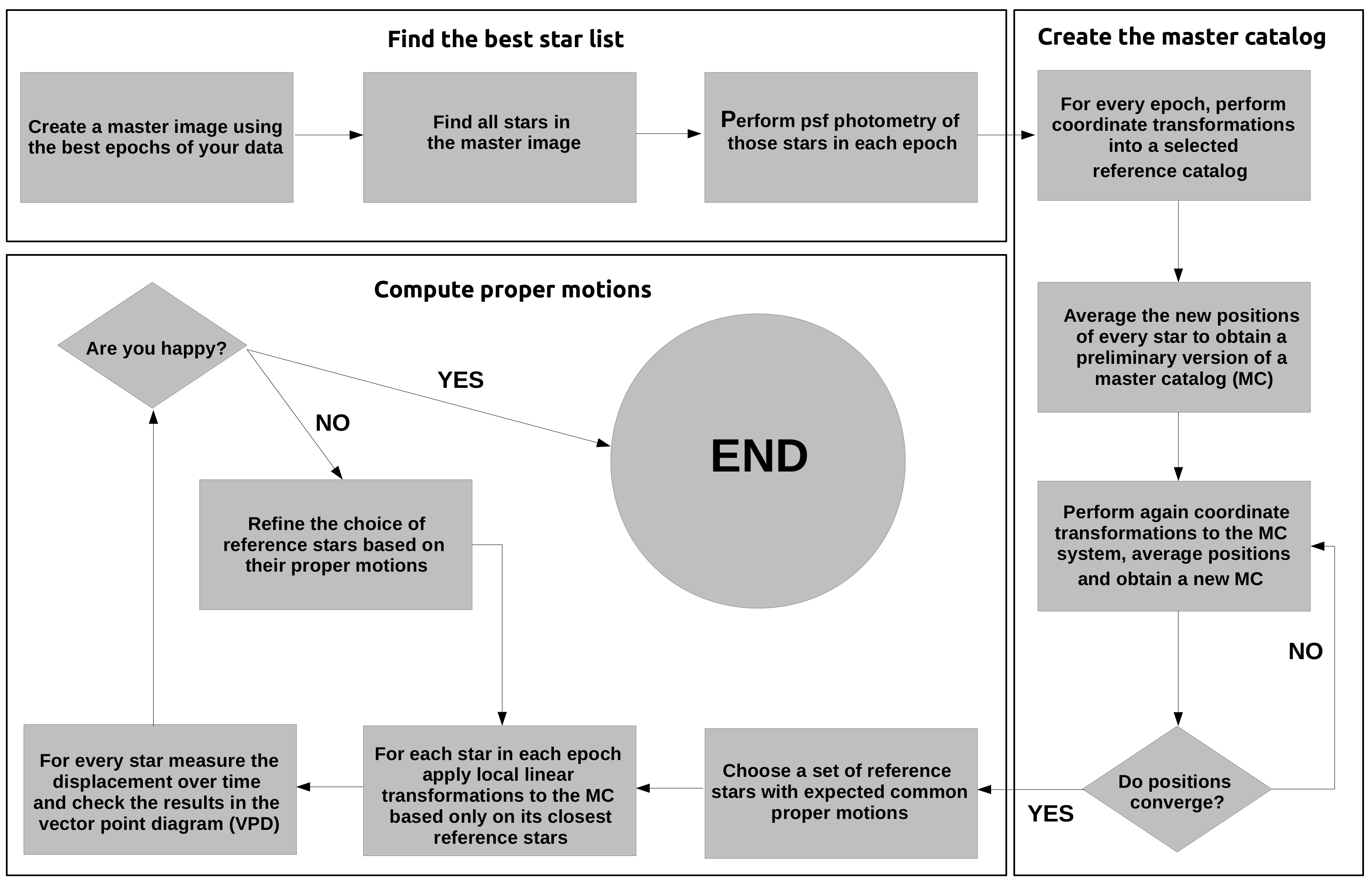}
\caption{Flow  chart  illustrating the  three  main  steps adopted  to
  computed PMs and explained in detail in the text.}
\label{flowchart}
\end{figure*}

\section{Proper motions}
\label{pm}

Relative PM studies  are often based on the comparison  of two sets of
images taken  in two  observing runs, separated  by a few  years.  The
data from each  observing runs are grouped and  considered as a single
epoch, and the PM is derived from the position difference of each star
between the  two epochs.  In the  present case, the data  from the VVV
survey extend almost continuously  across a time span of approximately
six years,  favoring the approach described  by \cite{bellini14}.  The
latter considers each image as a single epoch, and fits the PM of each
star, in  each of the x- and  y-axis, as the derivative  of the linear
relation  between the  star pixel  coordinate and  the time.   In this
section we  describe how this method  was optimized to  derive PMs for
stars  in  the  VVV  images,  applied  to a  field  including  the  GC
NGC~6544. Figure~\ref{flowchart} summarizes  the steps adopted to find
the  initial star  list, create  the master  (reference)  catalog, and
finally compute PMs. The details  of each of these steps are described
in the three subsections below.

\subsection{PSF photometry $\&$ the initial master list}
\label{sphoto_masterlist}

\begin{figure}
\includegraphics[width=8.5cm]{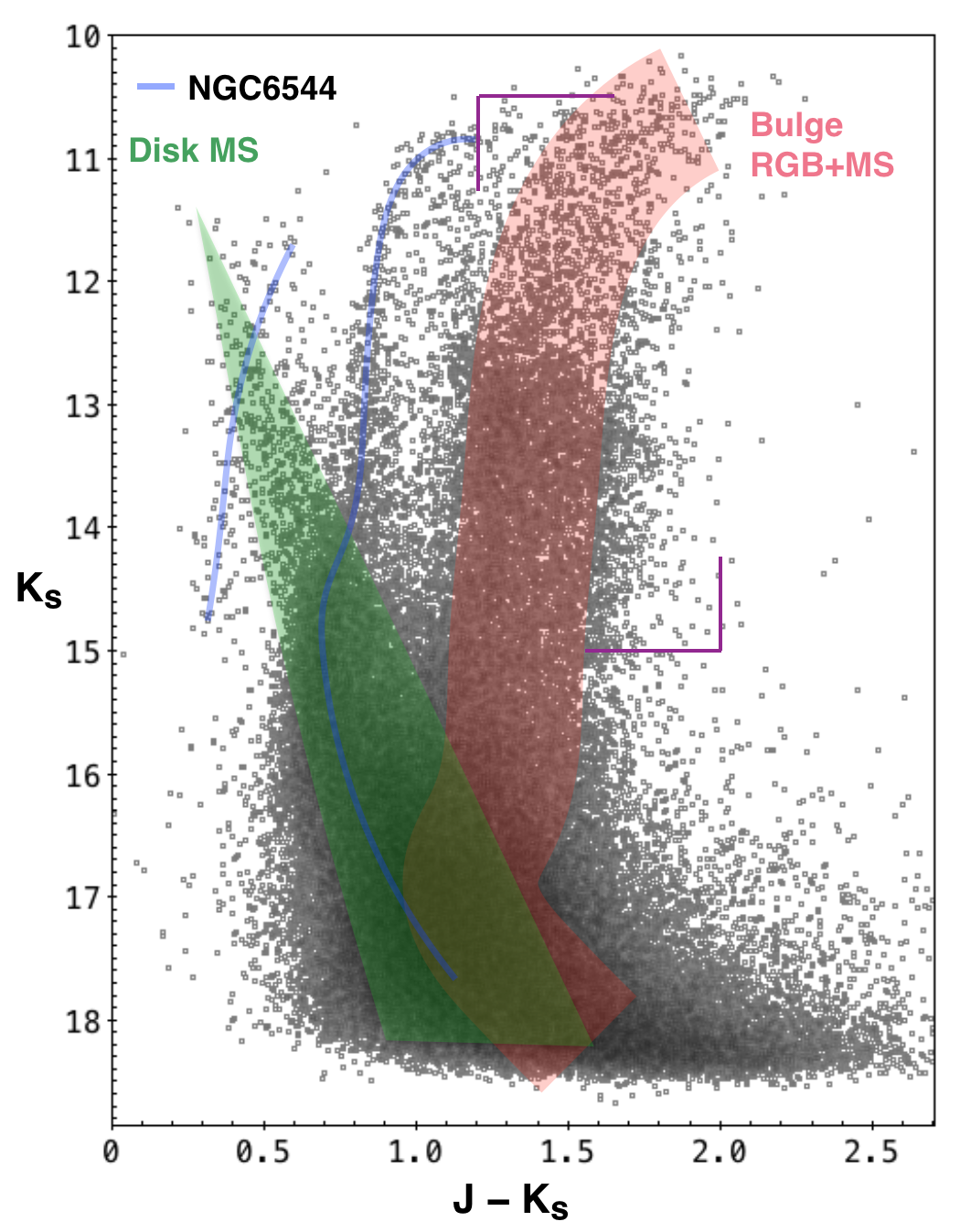}
\caption{CMD  for  all  the  stars  detected in  the  field  shown  in
  Fig.~\ref{FoV}.  The  hand drawn  blue  line  shows the  approximate
  location of  the cluster  sequences, while the  red and  green areas
  show the bulge RGB and disk MS, respectively.}
\label{cmdnoclean}
\end{figure}

The  data obtained  for the  VVV  survey comes  from the  VISTA/VIRCAM
instrument, an  array of $4\times4$ Raytheon VIRGO  detectors of $2048
\times  2048$  pixels each,  with  an  average  spatial resolution  of
$\sim$$0^{\prime\prime}$.339   pixel$^{-1}$  \citep{dalton06,emerson10}.
Each  individual science image,  called pawprint,  is composed  of two
exposures  dithered   by  40  pixels  in  order   to  remove  detector
artifacts. At  each sky pointing, six dithered  pawprints are acquired
in such a way  that they can be combined in a  single mosaic, called a
tile, filling  the large  gaps between the  16 VIRCAM  detectors.  All
pawprints are  processed through the  standard CASU pipeline  for bias
subtraction,       dark       correction       and       flat-fielding
\citep{emerson04,irwin04}.
For the present work,  PSF-fitting photometry was performed separately
on each given  detector of a given pawprint, and  the PMs were derived
from  the comparison  of all  the  epochs of  that specific  detector.
Other  nearby  detectors  were  ignored  here,  even  if,  because  of
dithering,  in different  pawprints they  cover part  of the  same sky
area.  For example, we initially  started the analysis on the detector
covering most of the cluster area. That was chip $\#15$ (out of 16) of
pawprint $\#2$  (out of 6) of  each observation available  for the VVV
field b309.  A $K_{s}$ image  of this chip is shown in Fig.~\ref{FoV}.
All  the available  epochs  for  chip $\#15$  of  pawprint $\#2$  were
downloaded from the ESO archive.   In each of the $ZYJHK_{s}$ filters,
we   retrieved  3,4,2,2,327   epochs,   respectively,  each   covering
$\sim$$12 \times 12$  square arcmin.  The  PM analysis  is based  on the
$K_{s}$  data  alone,  which  were  acquired between  April  2010  and
September 2015.
PSF  fitting  photometry on  each  image  was  carried out  using  the
DAOPHOTII/ALLSTAR  package \citep{stetson87},  with  a \cite{moffat69}
PSF model constructed using  the brightest, isolated and non saturated
stars  in  each image.   Initial  stellar  positions  and fluxes  were
measured independently on each image, with a relatively high detection
threshold  ($30\sigma$) and a  first run  of ALLSTAR  on all  of them.
This  ensured  that  only  stars  with good  photometric  quality  and
accurate positions were kept for the coordinate trasformation step. We
then  selected one  of  the images  with  best seeing  to  be used  as
reference   coordinate   system   ({\it   refframe})   and   run   the
DAOMATCH/DAOMASTER  \citep{stetson94}   codes  to  determine  accurate
frame-to-refframe  coordinate  transformations   for  each  image.   A
stacked mean image  was then created using the  MONTAGE routine of the
same package.  The latter has  high signal-to-noise (S/N) ratio, it is
free of cosmic rays, and it  has a field of view (FoV) slightly larger
than any  single epoch.  We will  refer to this stacked  mean image as
the master-image.  The star finding  (FIND) routine of DAOPHOT was run
on  the   master-image,  now   with  a  smaller   detection  threshold
($3\sigma$), together  with the aperture photometry  task (PHOT).  The
resulting master-list, after an inverse coordinate transformation, has
been used as  input for the ALLSTAR PSF-fitting  photometry routine on
each individual image, allowing the code to refine the position of the
centroids (ALLSTAR  parameter {\it RE = 1}).   A side advantage  of this
approach is that a unique ID number identifies a given star in all the
epochs.
The instrumental  magnitudes were calibrated to  the VISTA photometric
system  using $\sim$15~000  stars  in common  with  the public  CASU
catalogs derived for  the same epochs, by means  of a simple magnitude
shift.  The  final photometric catalog  was obtained by  averaging the
magnitudes of the individual  epochs.  It contains 140~000 stars, from
the  upper  red  giant   branch  (RGB)  down  to  approximately  three
magnitudes   below   the   main   sequence  turnoff,   as   shown   in
Fig.~\ref{cmdnoclean}.  The cluster sequences, on the blue side of the
CMD in  this figure,  are contaminated by  a well populated  bulge RGB
(parallel to the cluster RGB, but  to the red) and main sequence, both
with  a  large spread  in  magnitude  and  color due  to  differential
reddening  (and to  less  extent due  to  distance), and  by the  main
sequence of the  foreground disk, bluer and brighter  than the cluster
turnoff.  We note that the cluster's evolutionary sequences in the CMD
are less  broadened than the bulge  ones, suggesting that  most of the
reddening is behind the cluster.  The actual evidence from dust behind
the cluster is the ``great  dark lane'' that runs across low latitudes
toward the bulge discovered  by the VVV survey \citep{minniti14}.  The
cluster is located in the same direction of this great dark lane shown
in the left panel of Fig.~2 in \cite{minniti14}.


\subsection {The master catalog}
\label{smastercat}

The previous  steps yields a  photometric catalog for  each individual
epoch.  In  order to compare  the position of  each star in  all these
epochs, we need  to ensure that the pixel scale  is uniform across the
field of  view.  That  is because  a given star  may fall  on slightly
different positions  within the field,  on different epochs.   To this
end, the geometric distortion  correction specifically derived for the
VIRCAM camera in \cite{libralato15}  has been applied to each catalog.
This correction provides distortion-free coordinates with residuals of
the order of 0.023 pixels.

We  then  need to  define  a  master catalog,  to  use  as zero  point
reference in the computation of the displacement of each star, in each
epoch.  This was done by selecting $\sim$10~000 well measured stars in
common  among all  the epochs,  and use  them to  derive six-parameter
linear transformations  between each distortion-corrected  catalog and
the epoch  (refframe) previously used as reference  for the coordinate
transformations  described in  Sect.   \ref{sphoto_masterlist}.  These
transformations were then applied to each catalog, in order to obtain,
for each  star, N$_{\rm epochs}$ positions  that it would  have, if it
had  no PM,  in the  refframe system.   For each  star,  these N$_{\rm
  epochs}$ positions  were then averaged  to derive a  single $(X_{\rm
  ref},Y_{\rm ref})$  pair, that will  be the reference point  for the
calculation of  all the $\Delta  X=X_{\rm i}-X_{\rm ref}$  and $\Delta
Y=Y_{\rm i}-Y_{\rm ref}$, with $i=1,N_{\rm epochs}$.  The catalog with
the  reference positions  for all  the stars  is our  (initial) master
catalog.   The latter  is iteratively  refined, by  going back  to the
determination of  the six-parameter linear  transformations, this time
with  respect to  the  master  catalog instead  of  the refframe  one.
Convergence was typically achieved in less than five iterations.

It is  worth emphasizing  that the advantage  of creating  a reference
master catalog, instead of using a single epoch, is that the reference
catalog contains more  stars than any single epoch  catalogs.  At this
stage the coordinate transformations used to create the master catalog
are global for  the whole detector, rather than local  as they will be
in the  actual PM  calculation (see below).   This does  not introduce
more scatter  because the positions  in the master catalog  define the
zero point of the coordinate displacement of each star. Because the PM
will be the slope of the  $\Delta X$ (or $\Delta Y$) vs time relation,
if  the  zero points  are  slightly  different  for stars  located  in
different  regions  of  the  field,   this  does  not  affect  the  PM
calculation.

\begin{figure}
\includegraphics[width=8.5cm]{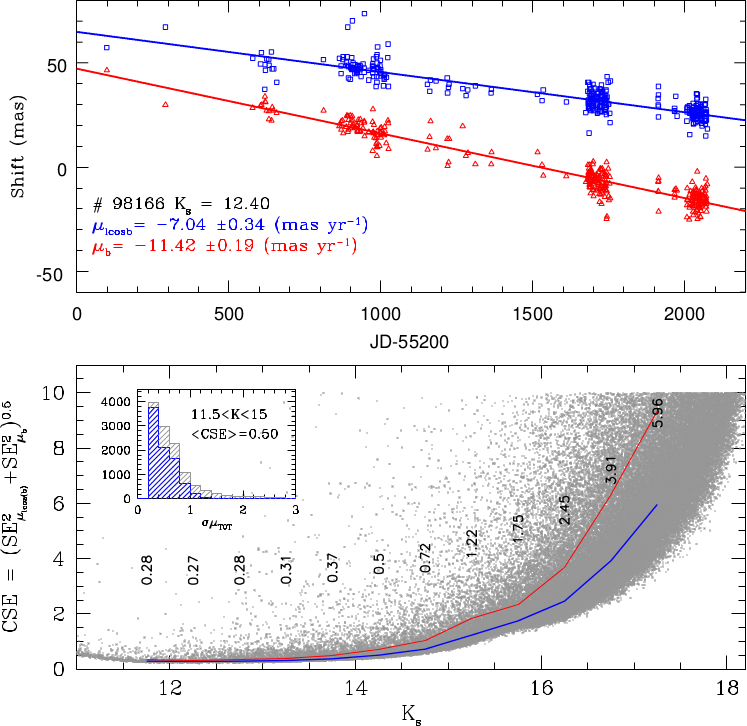}
\caption{ \textit{Top}:  Example of PM  determination by means  of the
  least square  fitting for one  of the cluster stars  (\#98166), with
  magnitude  $K_{s}=12.4$~mag.  Each  point  is the  star position  in
  different  epochs after  been locally-transformed  to  the reference
  system  of the  master  catalog.  \textit{Bottom}:  CSE  in the  PM,
  computed as the error on the slope in the linear fit shown above, as
  a  function  of  $K_{s}$  magnitude.   The thin  red  line  marks  a
  $3\sigma$ clipping on the error distribution; the thick blue line is
  the  median  error.   Numbers  list  mean  errors  in  bins  of  0.5
  magnitudes.   The figure  inset shows  the histogram  of  errors for
  11.5$<K_{s}<$15~mag  before (gray)  and after  (blue)  the $3\sigma$
  clipping.}
\label{shift_time}
\end{figure}

\subsection{Deriving proper motions}
\label{pm_derivation}

In order  to derive the PMs,  we will now transform  the coordinate of
every  star, in each  single epoch,  to the  coordinate system  of the
master catalog created above.  Then,  the displacement of each star in
each  epoch with  respect to  the master  catalog will  be calculated,
relative to the  displacement of a sample of  nearby stars selected as
reference  population.  In  other words,  we will  calculate  how much
star$N$ has moved,  compared to how much its  nearby stars have moved.
This relative displacement,  both in X and Y,  is then plotted against
time, and fitted with a linear relation whose slope (relative movement
per unit  time) is  by definition the  PM of  star$N$ along the  X and
Y-axis.

An example  of the  PM of a  star with magnitude  $K_{s}=12.40$~mag is
shown  in  the  upper  panel of  Fig.~\ref{shift_time}.   A  $3\sigma$
clipping has been  applied before the least square  fit.  The error on
the derived slope is, by definition, the statistical error (SE) on the
PM    in    each     axis.     The    combined    statistical    error
($\rm{CSE}=\sqrt{\rm{SE}^2_{\mu_{l}\cos(b)}+\rm{SE}^2_{\mu_{b}}}$), as
a  function  of $K_{s}$  magnitude  is shown  in  the  lower panel  of
Fig.~\ref{shift_time} for  all the stars.  The distribution  of the PM
CSE  at each given  magnitude is  obviously skewed,  with a  long tail
toward large PM errors.  
Two lines, representing the median  PM CSE for the unclipped stars and
a $3\sigma$ clipping of  the distribution are included for comparison.
The mean CSE  in bins of 0.5 magnitudes is  also indicated.  The inset
shows the histogram  of the PM CSE for  stars with $11.5<K_{s}<15$~mag
before and  after the  $3\sigma$ clipping.  The  mean CSE  within this
magnitude range is $0.50$ ~mas~yr$^{-1}$.

The  astrometric reference stars  selected for  the present  work were
bulge        RGB       stars.         In        previous       studies
\citep[e.g.,][]{bellini14,libralato15}  it has  been  common to  select
cluster  stars,   because  they  have  a   smaller  internal  velocity
dispersion, that is, they  move coherently.   Bulge stars  on  the other
hand, have a larger velocity  dispersion, therefore one needs a larger
sample  of them  in  order  to average  out  the different  individual
motions.  They have  been preferred in this work  because our ultimate
goal is to design a method that  can be applied to the whole VVV bulge
area, even far away from star clusters.

Well measured,  non saturated bulge RGB stars  were initially selected
based  on their location  in the  CMD (see  Fig.~\ref{cmdnoclean}) and
their  small photometric  error.   After the  first  iteration in  the
calculation  of PMs,  this initial  selection will  be  refined adding
their PMs as a selection  criterion. Iterations are repeated until the
number of selected reference stars converges.

Following \cite[][]{anderson06}, the  transformations used in the last
step of the  PM calculation are local, that is,  restricted to a small
region  of  the  chip  close  to  each star$N$.   This  is  more  time
consuming, but it ensures that residual astrometric distortions across
the field of view  of each chip do not have an  impact on our results.
Specifically,  in this  work the  displacement of  each star,  in each
epoch,  was  calculated  with  respect  to a  sample  of  the  closest
reference stars provided they are between 10 and 50 in number, and are
located within  a distance of 300  pixels from the  center of star$N$.
Stars  with less  than 10  reference stars  were not  analyzed.  These
nearby   reference  stars  were   used  to   derive  a   linear  local
transformation  between each  epoch and  the reference  catalog, valid
only  for star$N$. The  final PM  in pixels/yr  was then  converted to
~mas~yr$^{-1}$    by    means    of    the    VIRCAM    pixel    scale
(0$^{\prime\prime}$.339 pixel$^{-1}$).


\section{Kinematical decontamination of the cluster's CMD and error budget}
\label{sclaencmd}

\begin{figure*}
\centering
\includegraphics[scale=0.50]{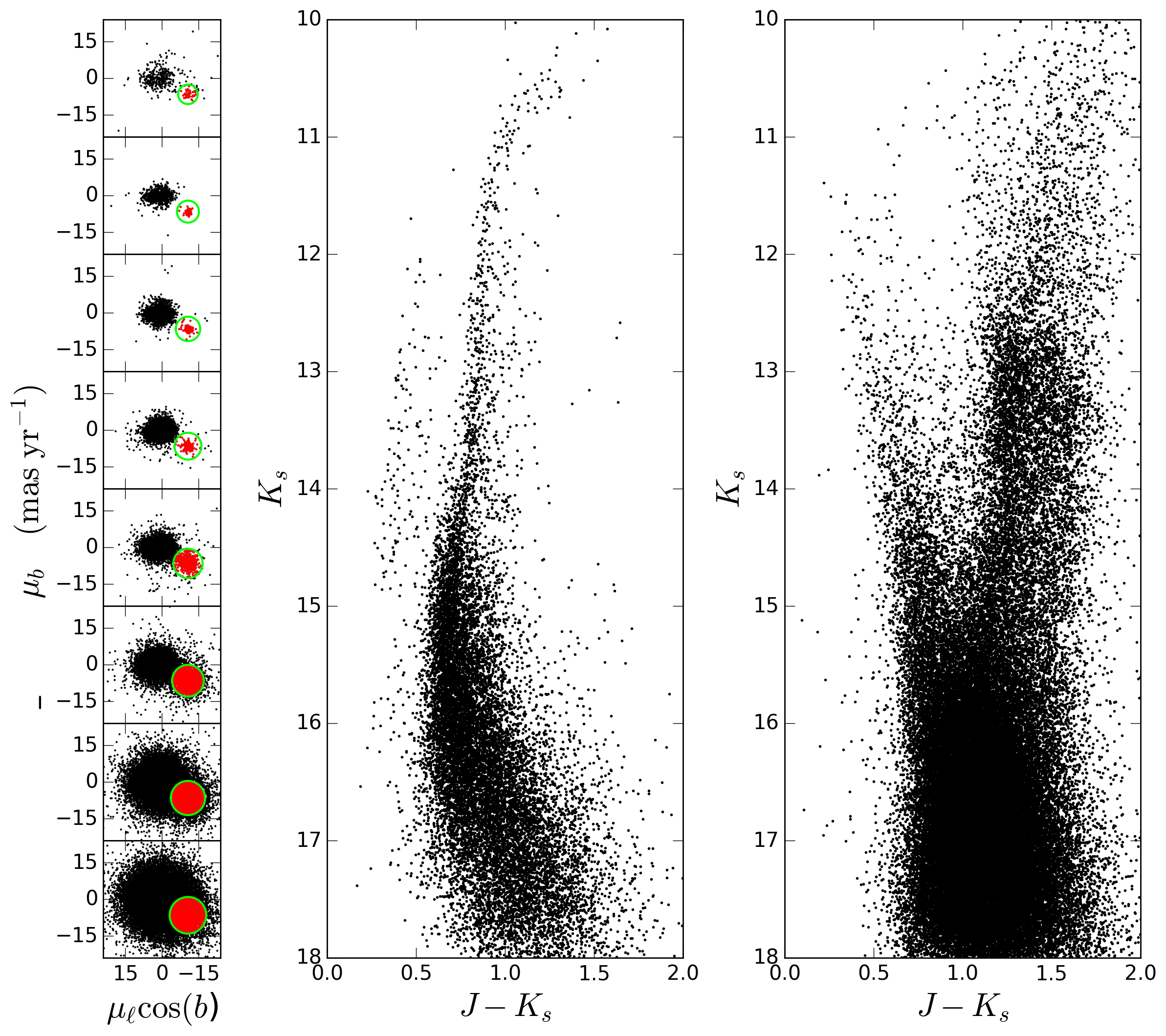}
\caption{  \textit{Left}:  Vector point  diagrams  for  all the  stars
  detected in the field, divided in eight one-magnitude bins according
  to  the  CMDs on  the  right. The  green  circle  shows the  adopted
  membership  criterion  used  to  selected cluster  members  in  each
  magnitude interval.  We  adopted a radius of 4  mas on the brightest
  bin and allowed a bigger  selection region for poorer measured stars
  (7.5  at the  bottom).   \textit{Middle}: clean  CMD  for the  stars
  assumed to be cluster members (red points inside the green circle in
  the  VPD).  \textit{Right}:  CMD  for the  field  counterpart  after
  excluding likely cluster stars.}
\label{bines}
\end{figure*}

The   main   result   of   our   PM   analysis   is   illustrated   in
Fig.~\ref{bines}. On the  left we show the Vector  Point Diagram (VPD)
of  all  the  stars  plotted  in  Fig.~\ref{cmdnoclean}  in  steps  of
magnitude, increasing  toward the bottom. We can  clearly separate (at
least) two  kinematically distinct  populations, a warmer  one roughly
centered  at  the  zero  of  the  VPD,  and  another  one  with  lower
dispersion,                         centered                        at
$(\mu_{l}\cos(b),\mu_b)\sim$$(-10.6,-6.6)$~mas~yr$^{-1}$.          By
construction, because  we choose bulge  stars as reference,  the cloud
centered  at (0,0)~mas~yr$^{-1}$  is made  of bulge  stars,  while the
tighter clump is made up of cluster stars.

Bona fide cluster  stars were selected as those  having PMs within the
small green circle,  centered in the tight clump.   The calculation of
the center  of this circle is explained  in Sect. \ref{svelocity}.  The
radius   was   arbitrarily  chosen   to   vary   smoothly  between   4
~mas~yr$^{-1}$ for the brightest  stars and 7.5 ~mas~yr$^{-1}$ for the
faintest ones.  Cluster stars selected  in this way, and marked in red
in   the  left   panels,  are   plotted   in  the   middle  panel   of
Fig.~\ref{bines}. All the other stars, marked in black in the VPD, are
mostly  field stars  and are  shown  in the  right panel  of the  same
figure.   As  expected,  cluster  stars lie  along  relatively  narrow
sequences  in the CMD,  showing a  blue horizontal  branch and  a well
defined main sequence turnoff at $K_{s}\sim$15.5~mag.  Field stars, on
the other hand, show the  usual bulge+disk CMD, with a prominent bulge
RGB,  a red  clump at  ($J-K_{s}$,$K_{s})\sim$(1.5,13)~mag,  both of
them widely  spread in color  due to differential extinction,  and the
disk main sequence on the blue side (see Fig.~\ref{cmdnoclean}).

The  PM distribution  of cluster  stars and  the SEs  obtained  in the
previous section allow  us to estimate the total  error of our method.
Indeed, the spread  of cluster stars in the VPD  is the convolution of
the real PM dispersion due to  the orbital motion of cluster stars and
the  total   (statistical  +  systematic)  PM   error.   The  observed
dispersion of  cluster stars  in the VPD  was calculated  by selecting
non-saturated  stars  with $K_{s}<15$~mag,  and  PM  CSE smaller  than
3$\sigma$,  resulting in  a  dispersion of  0.77 ~mas~yr$^{-1}$.   The
central velocity dispersion of NGC~6544  is not included in the Harris
catalog,  but  according to  the  spectroscopic  analysis  by Gran  et
al. (2017; in preparation)  seven confirmed NGC~6544 member stars have
a  radial  velocity   dispersion  $\sigma  \sim$6  ~km~s$^{-1}$.   The
analyzed stars are located at a mean distance of $3^{\prime}.52\pm2.9$
from the cluster  center, thus imposing a lower  limit for the central
velocity dispersion  of the cluster.   Adopting a cluster  distance of
2.5 kpc \citep{cohen14} this corresponds to an intrinsic PM dispersion
of 0.51 ~mas~yr$^{-1}$.  Subtracting in quadrature this value from the
observed  dispersion in  the  VPD, we  derive  a total  error of  0.57
~mas~yr$^{-1}$,  in  each  component,  valid for  cluster  stars  with
$K_{s}<15$~mag. As mentioned above, this  is the combination of the SE
plus the contribution of several possible sources of systematics, such
as   geometrical-distortion  residuals,  image   motion,  differential
chromatic  refraction,  S/N  bias,  local-transformation  bias,  among
others            \citep[see           e.g.,][and           references
  therein]{anderson06,bellini14,libralato15}.  The  CSE was derived in
Sec.~\ref{pm_derivation}  from the  slope  of the  relations shown  in
Fig.~\ref{shift_time}   for   all  the   stars,   resulting  in   0.50
mas~yr$^{-1}$.    The  total  error   estimated  above,   however  was
calculated for cluster stars, which are, on average, more crowded than
field stars,  and therefore  have slightly larger  SEs. Indeed,  if we
repeat  the calculation in  Fig.~\ref{shift_time} for  cluster members
only, we  get a  mean CSE of  0.62 ~mas~yr$^{-1}$.  By  subtracting in
quadrature  the SE ($\sim{\rm{CSE}/\sqrt{2}}$)  from the  total errors
estimated   above  (0.57   ~mas~yr$^{-1}$)  we   conclude   that  the
contribution   of   the   systematics   to   the   error   budget   is
0.37~mas~yr$^{-1}$  along each galactic  coordinate.  The  total error
for field stars  in each coordinate can now be  estimated by adding in
quadrature this systematic (0.37  mas~yr$^{-1}$) to the SE measured in
Sec.~\ref{pm_derivation} for disk and bulge stars separately (0.40 and
0.32  ~mas~yr$^{-1}$),  which   yields  0.54  and  0.50~mas~yr$^{-1}$,
respectively.

\section{The disk-bulge relative proper motion}
\label{spm_db}

\begin{figure}[t]
\centering
\includegraphics[width=8.5cm]{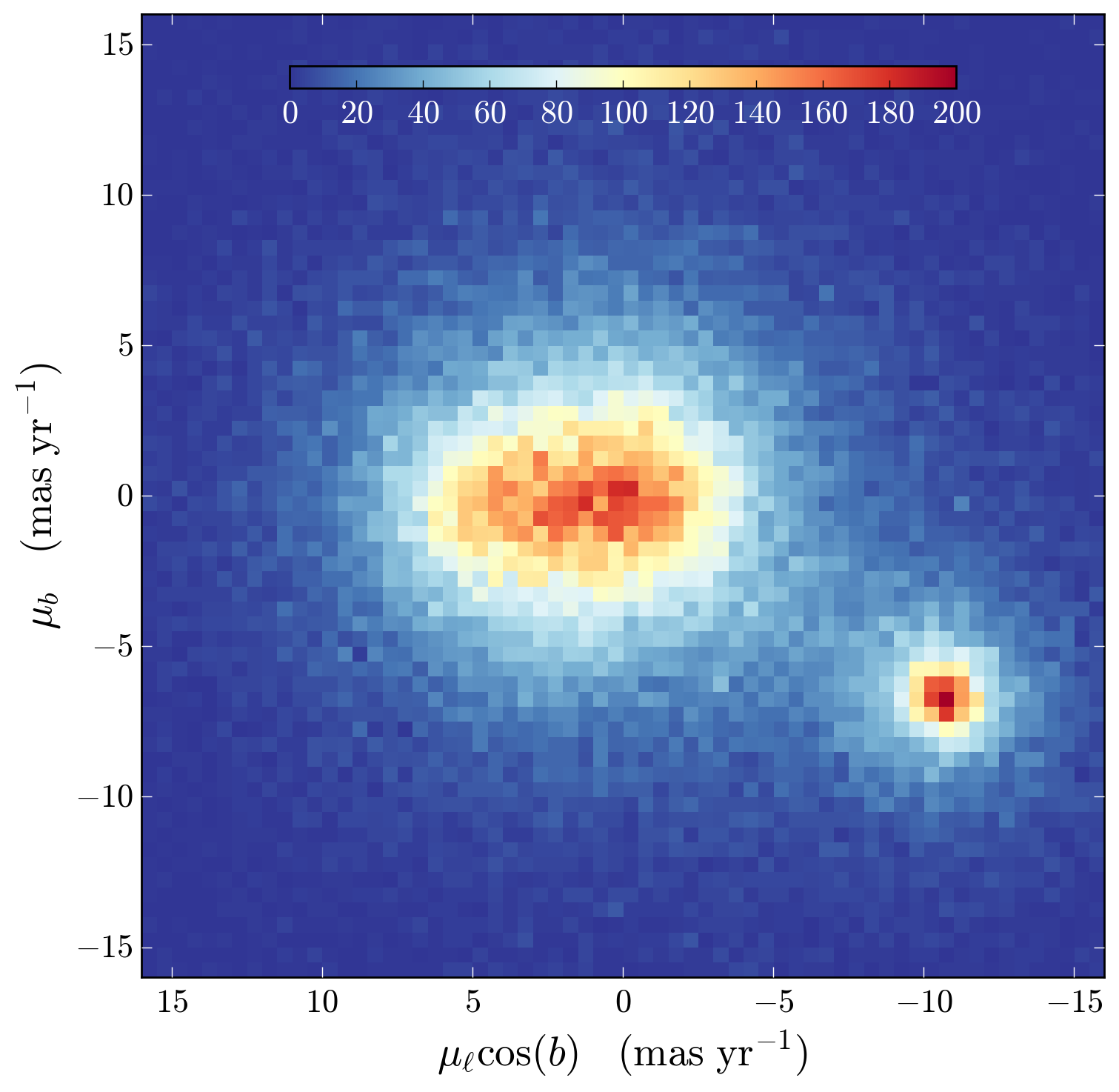}
\caption{Hess  diagram of  the VPD  for all the  stars with PM CSE  
lower than  10  ~mas~yr$^{-1}$. Stars  belonging to  NGC~6544
  cluster in  the lower right corner,  while field stars  are the cloud
  around  zero.  This  cloud  is  significantly  elongated  along  the
  longitude axis, because of the presence of two populations, disk and
  bulge, with the disk having a significant rotation.}
\label{vpdh}
\end{figure}

\begin{figure}
\includegraphics[width=9cm]{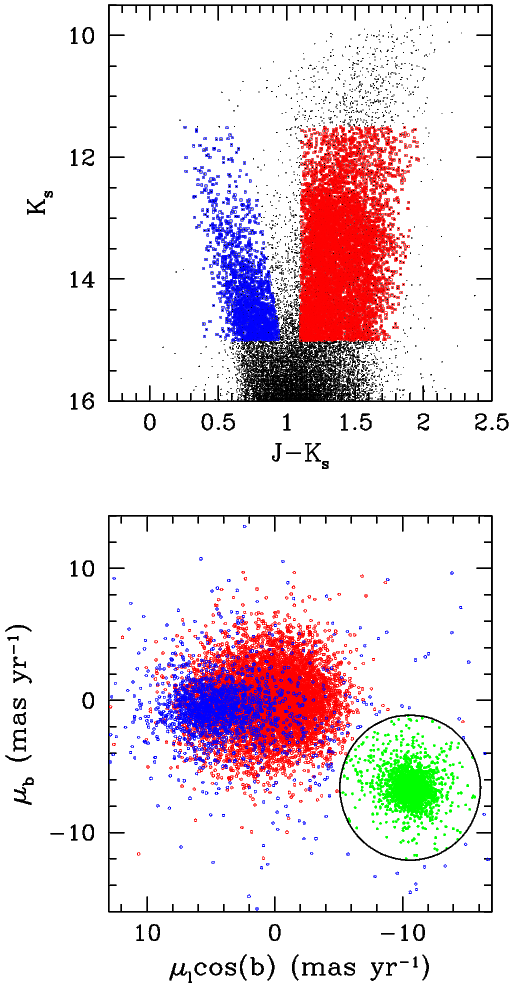}
\caption{\textit{Top}: CMD  showing our  selection of RGB  bulge stars
  (red)  and main sequence  disk stars  (blue) after  excluding likely
  cluster  members.   \textit{Bottom}:  VPD  showing the  relative  PM
  displacement between the selected stars from the CMD. NGC~6544 stars
  have been included in green for comparison.}
\label{pm_db}
\end{figure}

\begin{figure}
\centering
\includegraphics[scale=0.5]{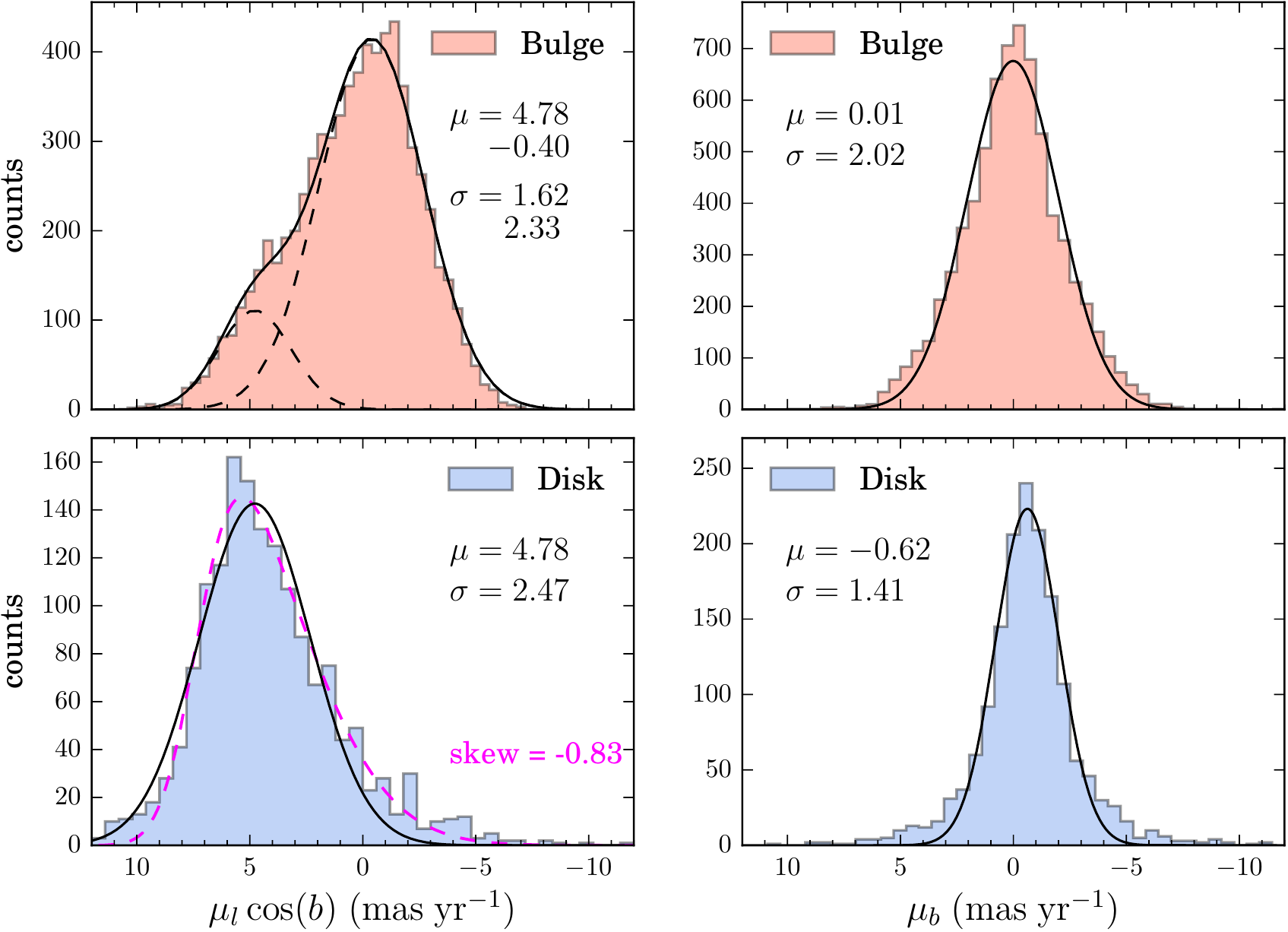}
\caption{PM distribution of bulge and  disk stars selected as shown in
  Fig.  7.  \textit{Left  panels}: $\mu{_l}\cos(b)$ distributions. The
  disk distribution is well fitted with a single Gaussian (black solid
  line) if the positive tail  is excluded. The otherwise better skewed
  Gaussian  fit  is shown  with  a  dashed  magenta line.   The  bulge
  distribution is fitted with two  Gaussians, with the centroid of the
  leftmost one  fixed to the  value found for the  disk. \textit{Right
    panels}: $\mu_b$  distributions of  bulge and disk  are symmetric,
  allowing a clean fit with  one Gaussian. In all panels, centroid and
  dispersion of the fitted profiles are quoted.}
\label{histograms}
\end{figure}

Figure~\ref{vpdh} shows the VPD for  all the stars with CSE lower than
10 ~mas  ~yr$^{-1}$ (see Fig.~\ref{shift_time})  color coded according
to  the density  of stars.   The PM  distribution for  field  stars is
highly elongated  along the longitude axis, suggesting  either a large
rotation or the presence of  two populations (or both).  In fact, from
the  Galactic position of  the cluster,  and from  the total  CMD, one
would expect the presence of
disk  stars in  front of  the cluster  and both  disk and  bulge stars
behind it.  In order to trace  the position of disk and bulge stars in
the VPD,  in Fig.~\ref{pm_db} (top panel)  we select a  sample of bona
fide disk stars, in the upper MS,  and a sample of bona fide bulge RGB
stars.   These two populations  partially overlap  in the  VPD (bottom
panel), although  disk stars are flatter  in this plane,  due to their
smaller  velocity  dispersion and  larger  rotation.   To compute  the
centroids and dispersion of both  components, we use the PM histograms
shown  in  Fig.~\ref{histograms} to  plot  the  bulge  and disk  stars
selected  in Fig.~\ref{pm_db}.   As can  be seen,  along  the Galactic
longitude axis  the PM  distribution of the  bulge (upper  left panel)
shows  a  secondary   bump  located  at  $\mu_{l}\cos(b)\sim$5.0  ~mas
~yr$^{-1}$, which coincides with the  peak of the disk distribution in
the same  direction (lower left  panel).  This feature is  produced by
disk stars in the RGB evolutionary sequence.  Taking this into account
the  procedure  was  the  following:  we started  analyzing  the  disk
component shown in the lower left panel of Fig. \ref{histograms} where
the best  fit was found to be  a skewed Gaussian, shown  with a dashed
magenta  line.   To  overcome  this asymmetry,  the  distribution  was
truncated  at  $\mu_{l}\cos(b)=2.0$ ~mas  ~yr$^{-1}$  which allowed  a
single  symmetric  Gaussian  centered  at  $\mu_{l}\cos(b)=4.78$  ~mas
~yr$^{-1}$ to be fitted.  Fixing  the disk centroid value to perform a
double Gaussian  fitting of the  bulge component along the  same axis,
the   centroid    of   this   component    was   found   to    be   at
$\mu_{l}\cos(b)\sim-$0.4  ~mas ~yr$^{-1}$.   The dotted  lines  in the
upper  left  panel  of  Fig.   \ref{histograms}  are  the  two  fitted
Gaussians whose  sum is the  solid line.  Along the  latitudinal axis,
both components  show a symmetric distribution, thus  allowing a clean
fit with one Gaussian each.
Notice  that  the exclusion  of  disk  stars  when fitting  the  bulge
component along the longitude axis  moved the latter slightly off from
the (0,0) position where reference stars are centered by construction.
The described procedure resulted in an offset of the disk with respect
to the  bulge RGB centroid,  by a $\Delta\mu_{l}\cos(b)=5.2  \pm 0.04$
~mas  ~yr$^{-1}$ and $\Delta\mu_{b}=-0.63  \pm 0.04$  ~mas ~yr$^{-1}$,
somewhat higher  when compared with  the relative velocities  found in
previous studies by \cite{clarkson08} and \cite{vasquez13}.  The bulge
RGB stars,  have an observed  dispersion of ($\sigma_l,\sigma_b)=(2.33
\pm 0.02, 2.02 \pm  0.02)$ ~mas ~yr$^{-1}$.  Deconvolving these values
from   the   total  PM   errors   estimated  in   Sec.~\ref{sclaencmd}
(err$_l=0.50$    ,   err$_b=0.50$)    ~mas   ~yr$^{-1}$    we   obtain
($\sigma_l,\sigma_b)=(2.28\pm0.02,   1.96\pm0.02)$   ~mas  ~yr$^{-1}$,
slightly smaller  than the  values obtained by  \citet[][]{soto14} and
\citet[][]{kuijken02}    but    in    very   good    agreement    with
\citet[][]{zoccali01}.      The    mild     PM     anisotropy    found
$\sigma_l/\sigma_b=1.16$  is in  perfect agreement  with \cite{zhao96}
and is explained as the effect of the bulge mean rotation.

\section{NGC 6544 space velocity}
\label{svelocity}

\begin{figure}[t]
\includegraphics[width = 8.5cm]{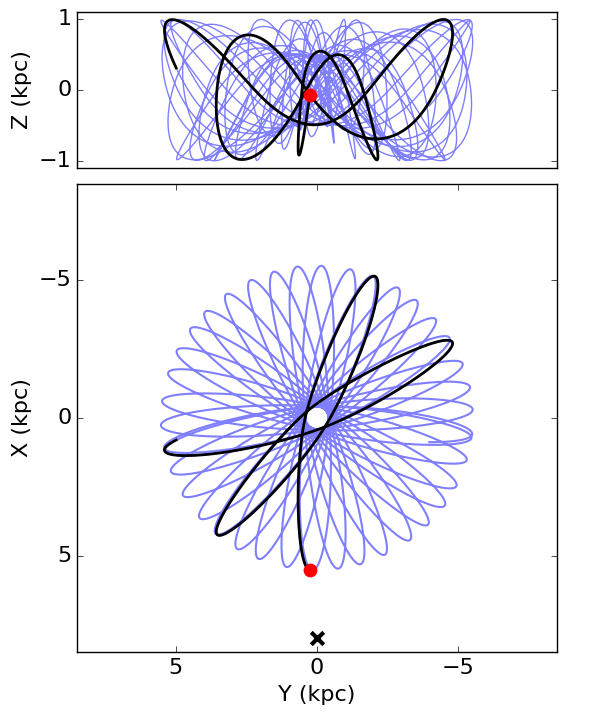} 
\caption{Computed orbit for NGC~6544 using an axisymmetric model.  The
  actual location of  the cluster is indicated by a  red dot while the
  position of the Sun by a  black cross. Blue and black paths show the
  orbit  of  the cluster  3  Gyr  back and  0.25  Gyr  ahead in  time,
  respectively.}
\label{sorbit}
\end{figure}

In order  to compute the space  motion of NGC~6544  we first determine
the  relative PM  of the  cluster with  respect to  the bulge,  as the
offset between their centroids in the VPD.  To this end, only clusters
stars with  $K_{s}<15$~mag and PM  CSE below the 3$\sigma$  limit (see
bottom  panel of  Fig.~\ref{shift_time}) were  selected,  to calculate
their mean PMs in both axis. The centroid of NGC~6544 turned out to be
at       ($\mu_{l}\cos(b),\mu_{b})=(-10.57\pm0.03,      -6.63\pm0.03$)
~mas~yr$^{-1}$.  Subtracting  the latter from  the value of  the bulge
centroid computed in the previous  section we obtain a relative motion
of    the    cluster    stars    with    respect    to    the    bulge
$\Delta\mu_{l}\cos(b)=-10.17\pm0.04$~mas~yr$^{-1}$                  and
$\Delta\mu_{b}=-6.62\pm0.04$~mas~yr$^{-1}$.
To determine  absolute PMs for  the cluster, we  have cross-correlated
the  cluster catalog  with  the UCAC5  catalog  recently published  by
\cite{zacharias17}  obtaining   330  stars  in   common.   From  their
comparison, we computed  a mean relative shift of  $-6.34\pm 0.39$ and
$-1.09\pm0.39$ ~mas~yr$^{-1}$  along galactic longitude  and latitude,
respectively.   The  derived  absolute  PM for  NGC~6544  resulted  in
($\mu_{l}\cos(b),\mu_{b})=(-16.91\pm0.39,-7.72\pm0.39$)
~mas~yr$^{-1}$.  In order to trace the orbit of the cluster, given the
current absolute PMs, we have used the program {\tt galpy} develped by
\citet[][]{bovy15}  adopting   for  NGC~6544  a   heliocentric  radial
velocity  of  $V_r=-27.3$  ~km~s$^{-1}$  from \citet{harris96}  and  a
distance d = 2.5 kpc from \citet{cohen14}.  Moreover, we have asumed a
Solar  motion  with   respect  to  the  Local  Standard   of  Rest  of
($U_{\odot}, V_{\odot}, W_{\odot})=(11.1, 12.24, 7.25)$ ~km~s$^{-1}$
from  \citet{schonrich10}, a  rotational velocity  of the  LSR  of 220
~km~s$^{-1}$ and a  distance of the Sun to the  Galactic center of 8.2
kpc \citep[][]{MWreview16}.   The calculation requires  the assumption
of  a  potential,  that  in  our case  includes  three  components,  a
spherical component (bulge) with  a power-law density distribution and
an exponential cut-off, a Miyamoto-Nagai disk, and a Navarro Frenk and
White halo  potential \citep{navarro97}.   The orbit has  been sampled
10~000~000  times  using  a  Runge-Kutta  6th integrator,  in  a  time
interval  corresponding to $-3$  Gyr to  $+250$ Myr.   Figure \ref{sorbit}
shows the orbit of NGC~6544 projected in the XY and RZ planes.  In our
integration, the  cluster reaches to  be closer than $\sim$0.5  kpc to
the   galactic   center.   As   discussed   by  \cite{minniti95}   and
\cite{bica16}, Galactic GCs can be divided in bulge, disk and halo GCs
according  to  their metallicity,  relative  position and  kinematics.
Taking into  account its relatively  low metallicity and based  on the
integration of this simple potential, we find that NGC~6544 is
more consistent with  the halo population than with  the bulge or disk
components.   It should be  noted, however,  that the  bulge potential
assumed    here    is   axisymmetric,    which    is   certainly    an
oversimplification.

\section{Membership probabilities} \label{membshipprob}

\begin{figure*}
\centering
\includegraphics[scale=0.7]{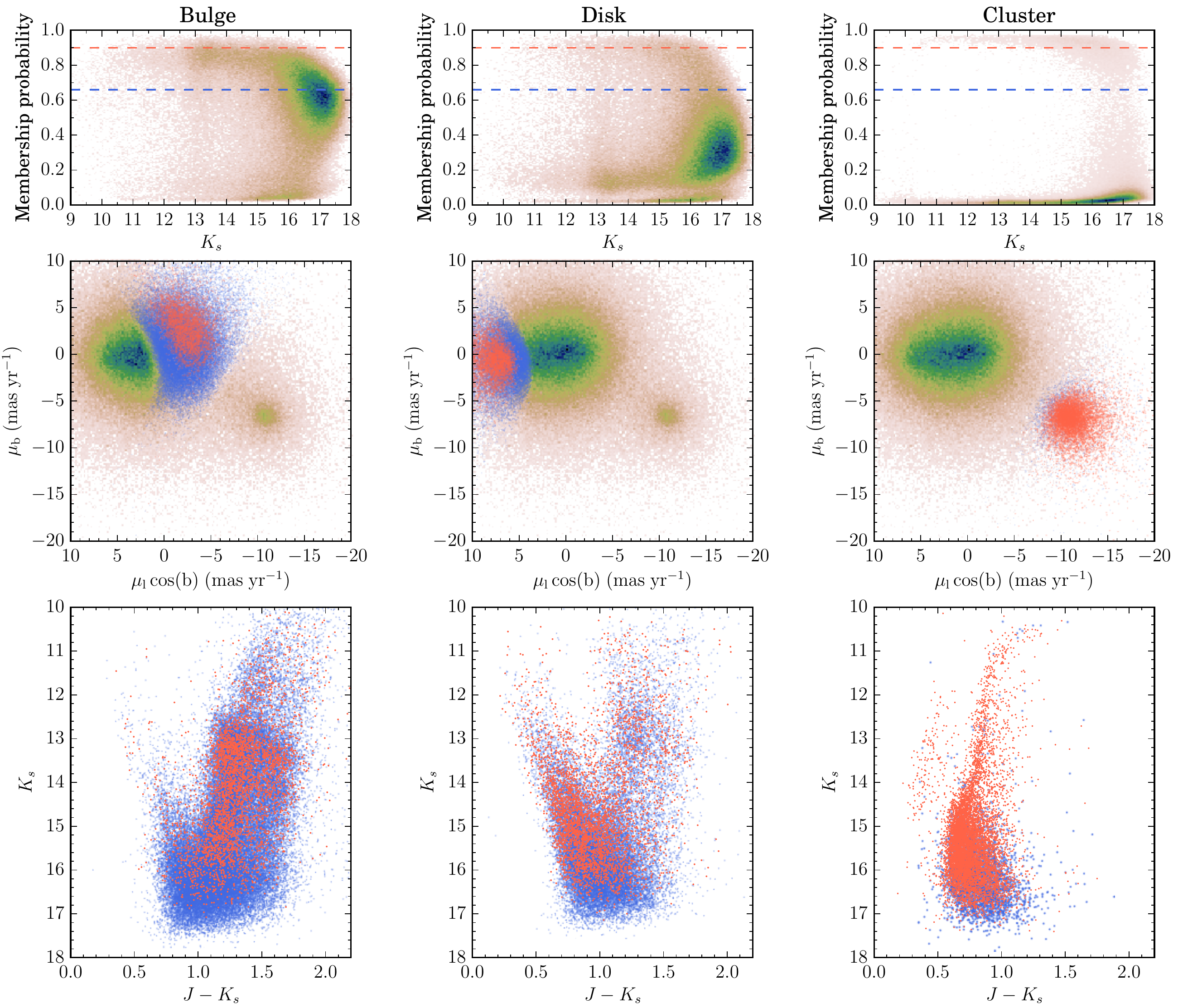}
\caption{\textit{Upper row:}  Membership probability as  a function of
  magnitude  for the  bulge, disk  and cluster,  color coded  by point
  density.  The blue  and red  dashed  lines depict  the $p=0.66$  and
  $p=0.90$  levels used  to select  likely members  of  the respective
  components of the mixture. \textit{Middle row:} The VPD of the whole
  sample  is  displayed identically  in  each  panel,  color coded  by
  density.  The blue  and red  points stand  for subsamples  of bulge,
  disk,  and   cluster  stars   selected  as  those   with  membership
  probabilities to  the respective component larger  than $p=0.66$ and
  $p=0.90$  (and  with  error  in  their  probabilities  smaller  than
  0.35). \textit{Lower row:} CMD for  the most probable members of the
  bulge, disk and cluster, color coded as above.}
\label{membership}
\end{figure*}

In order  to calculate the  likelihood of each  star to belong  to the
disk, bulge, or  GC we consider a probabilistic  approach based on the
Bayesian   partial  membership   (BPM)  model   for   continuous  data
\citep{heller08,gruhl14,blei07}.  Contrary to standard mixture models,
partial membership  models do not  restrict data samples to  belong to
only one of  the constituent distributions, but allows  data points to
have fractional  membership to multiple  components.  In the  BPM each
data  point is  modeled as  arising from  a geometric  average  of the
constituent distributions  weighed by a  membership probability vector
$\pi_i  \in  [0,  1]^K$  such  that  $~  \sum_{k=1}^K  \pi_{ik}  =  1$
\citep{heller08}. In  other words, each  data point has  an associated
membership  probability  vector  $\pi_i$   which  relates  it  to  the
components  of the  mixture.  This  approach allows  modeling  data in
scenarios with high overlap  between the generating distributions, for
example,  the samples  coming from  the disk  and the  bulge.   In the
present case we have a three-component  mixture model, that is, K = 3.
We  set the  components of  each population  as a  multivariate normal
distribution  $\text{MvN}(\mu_k, \Sigma_k)$,  where a  full covariance
and a  single variance matrix ($\Sigma  = I \sigma^2$)  are adopted in
the case of disk/bulge and  cluster, respectively.  In order to define
values for the mean vector  and dispersion of each MvN distribution in
the VPD, we  consider only bright stars ($K_{s}  \leq 15$~mag) with PM
CSE  $\leq3$ ~mas~yr$^{-1}$.   The centroids  of the  bulge,  disk and
cluster  are known and  kept fixed  at ($\mu_{xb},\mu_{yb})=(-0.40,0$)
~mas~yr$^{-1}$,  ($\mu_{xd},\mu_{yd})=(4.78,-0.62$) ~mas~yr$^{-1}$ and
($\mu_{xc},\mu_{yc})=(-10.57,-6.63$) ~mas~yr$^{-1}$, respectively.  In
the same  vein, initial values for  the scale parameter,  that is, the
diagonal of  the covariance matrix,  were set to $\sigma_b^2  = (2.28,
1.96)$, $\sigma_d^2 = (2.41, 1.30)$ and $\sigma_{gc}^2 = (0.51,0.51)$,
corresponding  to  the intrinsic  PM  dispersion (observed  dispersion
deconvolved  by  the  total   PM  errors)  of  each  population.   The
associated  errors  of  each  individual  PM  measurement  and  weakly
informative priors  were also included  in the covariance  matrices of
the components.  A detailed description of the  hierarchical model and
the specification of the priors can  be found in Appendix A.  In order
to compute  membership distributions, covariance  components and their
respective priors  we use an optimization approach  based on automatic
differentiation    variational   inference   \citep[ADVI,][]{jordan99,
  kucukelbir17}.  We  run the ADVI procedure  for approximately $1000$
iterations and  after convergence, new estimations of  mean values and
standard  deviations for  the  model parameters  were  drawn from  the
variational distribution. The described  procedure was run on a sample
containing 280~000  stars located in the inner  $12^{\prime}$ from the
cluster center  and with PMs errors lower  than 10~mas~yr$^{-1}$.  The
average  concentration of  the  bulge, disk  and  cluster found  after
convergence are $0.847$, $0.145$, $0.0074$, respectively.

Figure~\ref{membership} summarizes the  main results of this analysis.
The top panels show the average membership probabilities as a function
of $K_{s}$ magnitudes in the PM  space, for each of the three elements
of  the mixture.   The upper  right  panel shows  that likely  cluster
members, that  is, stars with more than  90$\%$ membership probability
to belong  to NGC~6544  separate clearly from  field stars down  to at
least $K_{s}\sim$16~mag.  For fainter  magnitudes the PM errors become
larger and  therefore many  stars have a  lower probability  to belong
either to the cluster or to  the field, that is, the two distributions
merge.  This  effect is even clearer  in the VPDs shown  in the middle
panels.  There  we highlight the stars with  probabilities higher that
$66\%$ and $90\%$  of belonging to the cluster  (right), disk (middle)
or  bulge  (left)  populations.   In  particular,  stars  with  90$\%$
probability of being cluster members set aside from field stars, while
for probabilities  lower than $\sim$$70\%$ the cluster  blob merges with
the  bulge+disk one  and  the contamination  from  the field  steadily
grows.  The  bottom panels of  Fig.~\ref{membership} show the  CMD for
bulge, disk  and cluster stars obtained  using the same  color code as
the middle panels.  As can be seen, stars with a 90$\%$ probability of
being cluster  members can be found  even $\sim$2 mag  below the main
sequence  turnoff.   This  kind  of  quantitative  assessment  of  the
probability membership  of each star will be  particularly relevant in
the context  of future spectroscopic  follow-ups of cluster  (or bulge
and/or disk) stars.

\section{Metallicity, interstellar reddening and age}

A major benefit of the  membership probability analysis is that we can
now separate the cluster population from the field stars and produce a
CMD  with  the most  likely  cluster  members.  However,  differential
reddening can  still produce a  significant width on  the evolutionary
sequences of the cleaned CMD. To correct for this effect, a relatively
extended  field centered in  NGC~6544 was  divided in  64, $2^{\prime}
\times2^{\prime}$ square subfields. One  of them with well defined and
populated  sequences was  selected to  be the  reference  subfield.  A
fiducial line  was extracted from it  and used to  apply the technique
described  in \cite{piotto99}.  Namely, the  CMD of  each  subfield is
shifted along the  reddening vector to match the  fiducial line of the
reference  CMD. In  this  manner  we obtained  the  CMD corrected  for
differential reddening of each single subfield.
\begin{figure}
\centering
\includegraphics[width=8.5cm]{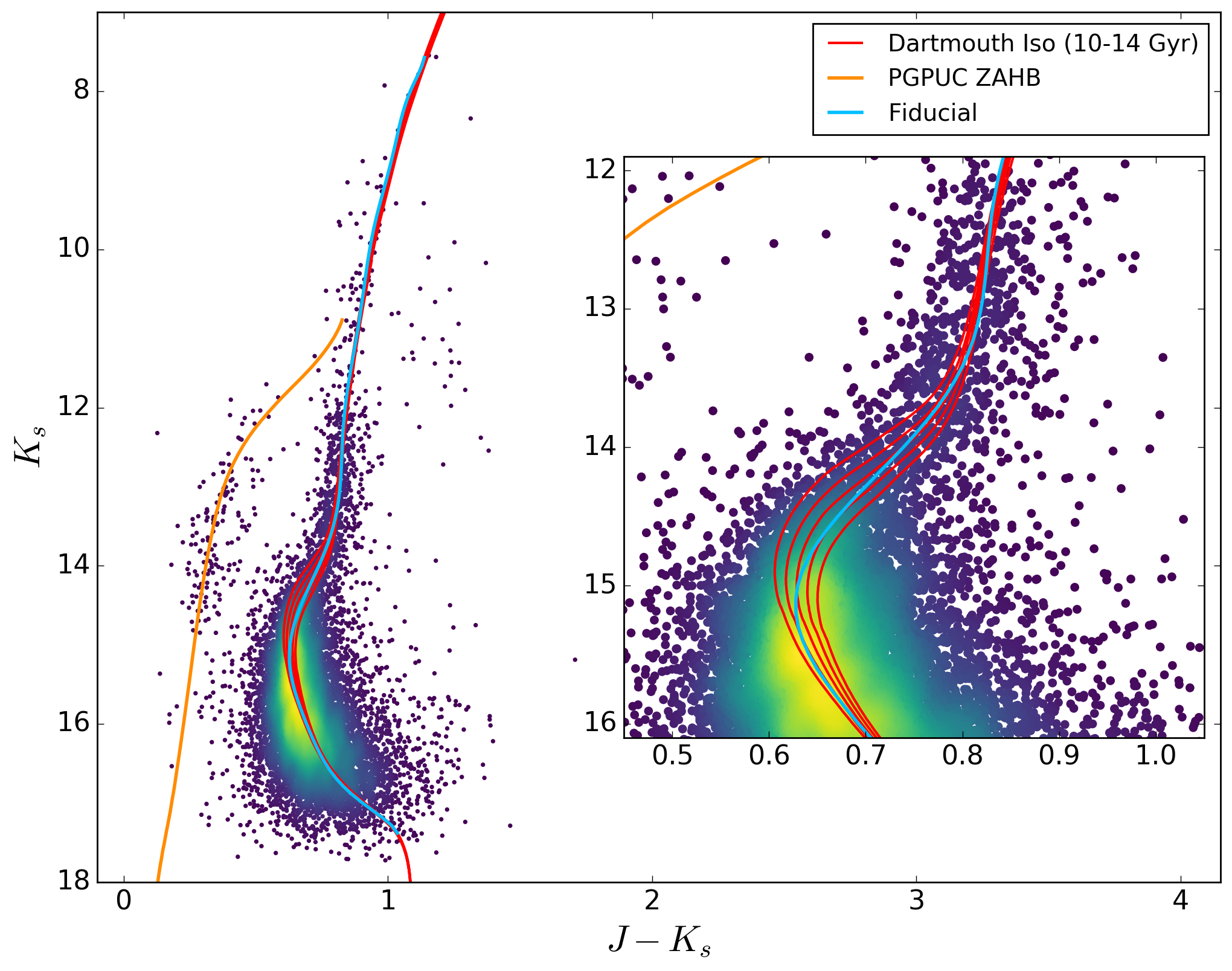}
\caption{CMD  including only  those stars  with $P_c(i)$  greater than
  70$\%$ of being cluster members.  The blue solid lines represent the
  ridge line obtained using VVV data. We adopted a distance modulus of
  $DM_0$  = 11.96  from \cite{cohen14}  to match  the  isochrones from
  Dartmouth (red) and PGPUC  (yellow) databases. The inset shows that,
  at the level of the SGB-MS,  the best fit of the isochrones with the
  fiducial line is obtained for an age between 11 and 13 Gyr.}
\label{cmdisocrona}
\end{figure}
Selecting only  stars with  membership probability higher  than $70\%$
from the  reddening corrected CMD, the main  evolutionary sequences of
NGC~6544  can be  easily traced  and can  be used  to  derive reliable
cluster  parameters as  shown in  Fig.~\ref{cmdisocrona}. Most  of the
bright stars in  this CMD ($K_{s}<12.0$ mag) are  saturated in the VVV
catalogues. Accordingly,  they were included here  from the photometry
by \cite{valenti10}, transformed from  the 2MASS $JHK_s$ system to the
VISTA $JHK_s$ system  using the most up to  date color transformations
from Gonzalez-Fernandez  et al. (2017; in  preparation).  To determine
the metallicity  and reddening of  the cluster, the fiducial  line was
used to match at the RGB  level a set of isochrones from the Dartmouth
stellar  evolution database  \citep{dotter08},  adopting the  distance
modulus  derived  by  \cite{cohen14}  ($DM_0=11.96$  mag).   The  best
coincidence  was obtained for  values of  $-1.5$ dex  and 0.36  mag in
metallicity  and  reddening respectively.   We  note  that across  the
studied area,  the computed reddening varies between  0.34 to 0.66~mag
and therefore the  value given here applies only  to the area selected
to  constructed  the fiducial  line  of  the  cluster.  We  also  have
included  a  zero-age  horizontal  branch  isochrone  from  the  PGPUC
database \citep{valcarce12} transformed to the filters used in the VVV
survey.  We have  excluded the hydrogen burning locus  in the analysis
since  these isochrones  do not  include  He diffusion  along the  RGB
phase, which  in this  particular case produces  a clear shift  to the
blue when  comparing with isochrones  from the Dartmouth  database for
the same  adopted distance and  metallicity values.  The fit  with the
horizontal  branch  population  confirms  that  the  adopted  distance
modulus  is correct  within  the observational  errors.   In order  to
estimate the age  of NGC 6544, we visually  compared the fiducial line
with the five  isochrones that best match our  data, representing ages
from 10 to  14 Gyr respectively (see Fig.  \ref{cmdisocrona}).  As the
inset  of  Fig. \ref{cmdisocrona}  shows,  the  fiducial  line of  the
cluster  falls down  along  the  sub giant  branch  and main  sequence
intersecting the  three innermost isochrones, suggesting  that the age
of NGC~6544 lies between 11$-$13 Gy.

\section{Elongation of NGC~6544}
\label{sextension}

\begin{figure*}[!]
\centering
\includegraphics[width=18cm]{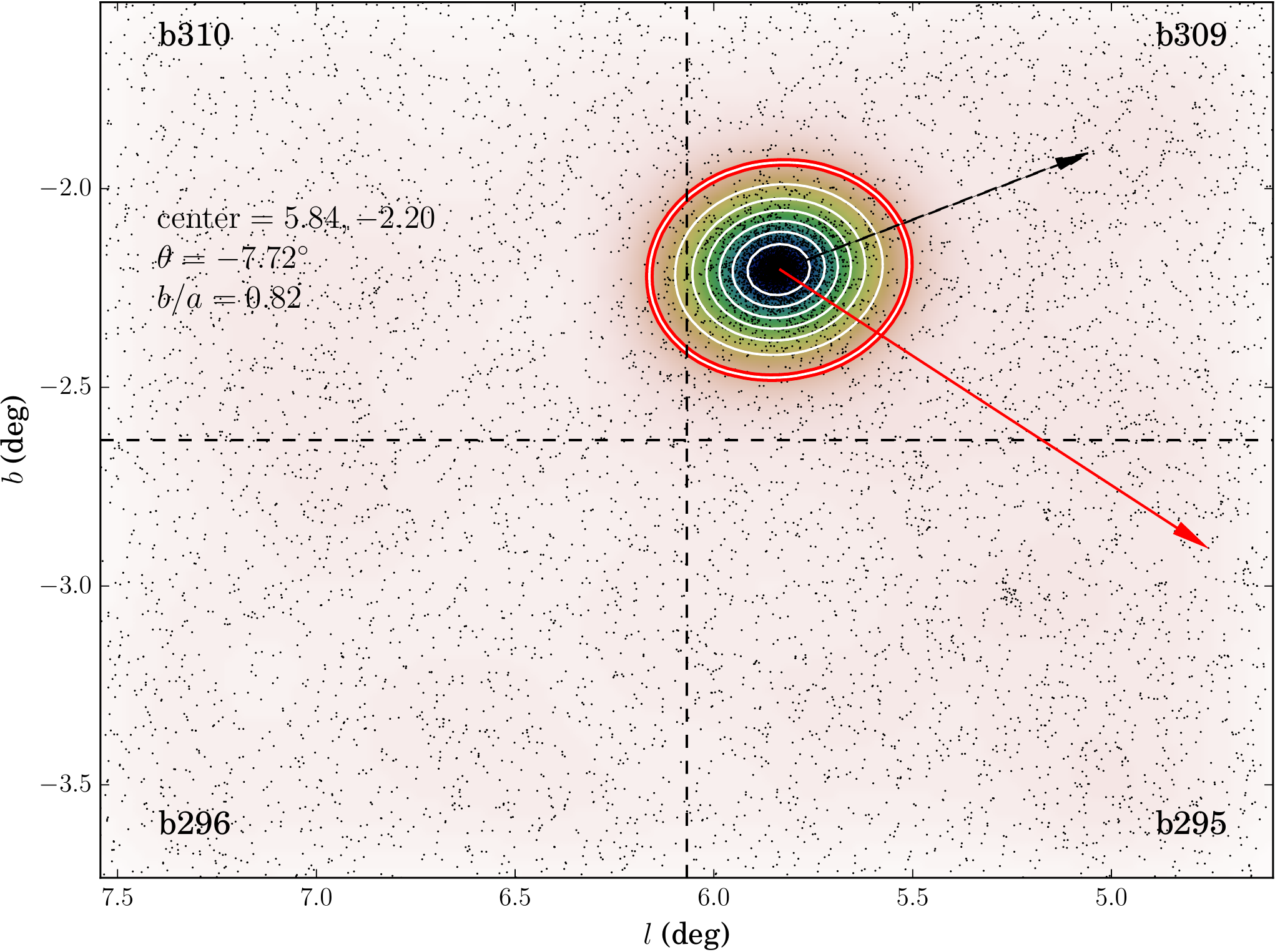}
\caption{Galactic  coordinate map  of the  area covered  by  the tiles
  b295,  b296, b309  and b310.   Black points  depict the  position of
  likely  cluster stars  with low  PM errors.  A color  map highlights
  their surface  count density,  and a set  of contour lines  is drawn
  around  the center  of NGC~6544.   The black  and red  arrows points
  toward the Galactic  center and along the direction  of the cluster
  orbit,  respectively.  The   cluster  center,  eccentricity  of  the
  outermost contour line, and the angle between its major axis and the
  Galactic equator are quoted.}
\label{mapa}
\end{figure*}

Based on the  cluster rather flat radial density  profile and inverted
mass function, \cite{cohen14} suggested that it might be more extended
than  previously  thought,  loosing  stars due  to  the  gravitational
interaction with the Galaxy.  PMs can be very helpful to trace cluster
members well outside the tidal  radius, hence we decided to extend our
analysis to  a region  $\sim$$3^{\circ} \times 2^{\circ}$  wide across
the  cluster.  Specifically,  we  applied the  procedure described  in
Sec.~\ref{pm}  to tiles  b295, b296,  b309  and b310,  compiling a  PM
catalog for more than 15  million stars. In order to identify probable
cluster  members, we  have selected  stars  with PM  CSE smaller  than
1.5~mas~yr$^{-1}$  ($K_{s}$ $\lsim$ 15.5~mag)  and within  a  radius of
2~mas~yr$^{-1}$  centered  at  ($-10.57,  -6.63$)  in  the  VPD.   The
resulting sample includes 11~217  stars, whose spatial distribution is
shown  in Fig.~\ref{mapa}.   The map  shown here,  at the  distance of
NGC~6544, corresponds  to an area of $129\times96$  pc$^{2}$. Far from
the cluster  center, the  map shows a  rather uniform  distribution of
stars, evidence  that a  small contamination from  the field  is still
present inside the ``cluster'' zone  of the VPD.  If tidal tails would
be present, however, we would expect an enhancement of this density in
the direction of the cluster  orbit, which is not seen.  Obviously, we
are considering here only stars  with mass slightly smaller than about
0.8 $M_\odot$ ($K_{s}$ $\lsim$ 15.5~mag). The present data do not allow
us to rule out the possibility that, due to dynamical segregation, the
stars that were tidally stripped were only the low mass ones.

In Fig.~\ref{mapa} we also  show the surface stellar density, obtained
via a Gaussian  kernel estimation (with a kernel  bandwidth set by the
custom  Scott's  rule).  The  spatial  distribution  of cluster  stars
across the  inner $\sim$12 arcmin  from the cluster center  is clearly
elongated.  This shape is accentuated by a set of contour lines, which
were  used to  estimate an  axial ratio  b/a $=0.82$. According  to the
elongation  values presented  by \cite{chen10}  for 116  Galactic GCs,
NGC~6544 is  among the $30\%$ most  elongated clusters. Interestingly,
the  cluster elongates roughly  in the  direction pointing  toward the
Galactic center.  The origin  of this morphological signature might be
related with  the gravitational  potential of the  MW causing  a tidal
perturbation  in the direction  of the  Galactic center,  an encounter
with the Galactic plane or  by a relatively high rotation of NGC~6554.
However, according to \cite{chen10}, those  GCs close to the bulge and
exhibiting obvious  flattening tend to have  their elongation pointing
toward the Galactic  center due to tidal effect  from the bulge.  This
is  in  very  good  agreement  with  the  results  obtained  in  Sect.
\ref{svelocity}, where  the simulated orbit locates  NGC~6544 as close
as 0.5 Kpc from the Galactic center.

\section{Summary \& conclusions}
\label{sconclusions}

In  this paper  we described  the procedures  adopted by  our  team to
compute PMs using VVV data.   In particular we have followed the local
coordinate transformations approach described by \cite{anderson06} and
\cite{bellini14}. In order to validate  the method, we have applied it
to a region $\sim$$3^{\circ} \times 2^{\circ}$ wide (4 tiles), toward
the MW bulge, including the GC NGC~6544. The six-year time baseline of
$K_{s}$-band  observations provided by  the VVV  survey allowed  us to
achieve  a   typical  accuracy  in  each  PM   coordinate  lower  than
0.51~mas~yr$^{-1}$  (statistical+systematics)   ,  for  $K_{s}<15$~mag
gradually increasing  up to 10~mas~yr$^{-1}$ at $K_{s}  = 18$~mag. The
PMs allowed us to establish  solid membership for cluster stars, which
separate very well from field stars  in the VPD, and to derive for the
first   time,  the   absolute  PM   of  this   system,   resulting  in
$(\mu_{l}\cos(b),\mu_b)=(-16.91\pm0.39,-7.72\pm0.39)$ ~mas~yr$^{-1}$.

Based on the  computed PMs and spatial distribution,  we applied a BPM
model  to select  the  most  likely cluster  members,  and produced  a
decontaminated CMD, essentially free  of field stars. Additionally, we
correct for  differencial reddening, allowing us to  better define the
evolutionary sequences in  the CMD.  From the comparison  of the clean
and differential reddening corrected  cluster CMD with isochrones from
Dartmouth database, the metallicity  and reddening of the cluster were
derived,   finding  values   in   good  agreement   with  those   from
literature. Additionally,  an age of $\sim$11$-$13  Gyr was estimated
for  NGC~6544.  Combining  PMs and  the  values for  the distance  and
radial velocity taken from  literature, we computed the Galactic orbit
of   this  cluster  assuming   an  axisymmetric   model  for   the  MW
gravitational  potential.   Based on  this  simplified assumption,  we
concluded  that NGC~6544  is  likely  a GC  associated  with the  halo
component  of  our  Galaxy.  This   is  in  accordance  with  its  low
metallicity  (Fe/H $\sim-1.5$ dex),  given that  bulge  clusters are
usually relatively more metal rich \citep{minniti95,bica16}.

The PM distribution of both disk and bulge stars were also obtained in
this  work.   The  relative  PM  between  the  two  was  found  to  be
($\Delta\mu_{l}\cos(b),\Delta\mu_{b}$)  $= (5.2\pm0.04, -0.63\pm0.04$)
~mas~yr$^{-1}$, while the PM dispersion of bulge and disk stars, along
each       galactic      axis       was      measured       to      be
$\sigma_{\mu_l\cos(b)}=2.28\pm0.02$,  $\sigma_{\mu_b}=1.96\pm0.02$ and
$\sigma_{\mu_l\cos(b)}=2.41\pm0.04$,       $\sigma_{\mu_b}=1.30\pm0.04$
~mas~yr$^{-1}$,  respectively.  The slightly  higher PM  dispersion of
the bulge  along the longitude axis  would be due to  its rotation, in
agreement with previous studies.

The extended spatial coverage of this study allowed us to look for the
footprints  of  the  interaction  between  NGC~6544  and  the  Galaxy.
Although no  signatures of  extra tidal stars  have been found  in the
distribution  of   cluster  stars,  we  detected   a  relatively  high
elongation  of the  cluster in  the direction  of the  Galactic center,
possibly  associated  to  the  tidal  forces exerted  by  the  Galaxy.
Membership probabilities computed in this  work will be of great value
for future  spectroscopic follow-up studies of cluster  stars, as well
as the  bulge and  disk.  The present  study shows that,  although not
optimized for this purpose, VVV  data are suitable for high-quality PM
studies.   The VVV  eXtended (VVVX)  Survey  is now  ongoing and  will
increase the time baseline by three more years, thus further improving
the PM accuracy.

\begin{acknowledgements}
We  gratefully  acknowledge  support   by  the  Ministry  of  Economy,
Development, and Tourism's Millennium Science Initiative through grant
IC120009, awarded  to The Millennium Institute  of Astrophysics (MAS),
by  FONDECYT Regular  1150345, 1170121,  1170305, 1171678  and  by the
BASAL-CATA Center for Astrophysics and Associated Technologies PFB-06.
F.   G.   acknowledge  support  from CONICYT-PCHA  Doctorado  Nacional
2017-21171485.   M.   G.   acknowledges  financial  support  from  the
Millennium  Nucleus RC130007  (Chilean Ministry  of Economy)  and from
FONDECYT grant 1141175. A.A.R.V. acknowledges support through FONDECYT
postdoctoral grant 3140575. We warmly thank Andrea Bellini for sharing
with us his  experience in astrometry, at the  beginning of this work.
We also thank the referee  for its comments which greatly improved the
present manuscript.

\end{acknowledgements}

\appendix
\label{appendixA}
\section{Bayesian mixture model for partial clustering} \label{appendix-bayesian}

In  this  section we  describe  the  Bayesian  hierarchical model  for
partial membership \citep{heller08,gruhl14,blei07} of PM for a mixture
of   $K=3$   multivariate   normal  distributions   $\text{MvN}(\mu_k,
\Sigma_k)$.   Let  $y_i   \in  \mathbb{R}^D$   be   the  observations,
$i=1,\ldots,   N$   with    associated   uncertainties   $\Sigma_i   =
\text{diag}(\epsilon_{xi}^2,  \epsilon_{yi}^2) \in \mathbb{R}^{D\times
  D}$,  where $\epsilon_i$ corresponds  to the  individual measurement
errors.  Each  observation has  an  associated membership  probability
vector $\pi_i \in  [0,1]^K$ which relates it to  the components of the
mixture. Conditioned to the membership probability vector we have

\begin{equation} \label{eq-data-gen}
    x_i | \pi_i \sim \text{MvN} \left(\widehat \mu_i, \widehat \Sigma_i \right),
\end{equation}
where
\begin{equation} \label{eq-data-Smu}
    \widehat \Sigma_i =  \Sigma_i + \left(\sum_{k=1}^K \pi_{ik} \Sigma_k^{-1} \right)^{-1}  
\end{equation}
and
\begin{equation} \label{eq-data-Smu}
    \widehat \mu_i = \left(\sum_{k=1}^K \pi_{ik} \Sigma_k^{-1} \right)^{-1} \left( \sum_{k=1}^K \pi_{ik} \Sigma_k^{-1} \mu_k \right).
\end{equation}

We note that  each $x_i$ is generated by  a single multivariate normal
distribution, where  the natural  parameters are weighted  averages of
the  parameters  of  the   individual  distributions  of  the  mixture
\citep{heller08}. The membership probabilities $\pi_i$ are constrained
to  the  simplex\footnote{Each  individual  component  of  $\pi_i$  is
  positive and $\sum_{k=1}^K \pi_{ik} = 1$}, as done in \cite{gruhl14}
using the logit or softmax function

\begin{equation} \label{softmax}
    \pi_{ik} = \frac{\exp(\eta_{ik})}{ \sum_{j=1}^K \exp(\eta_{ij}) },
\end{equation}
where $\eta_i \in \mathbb{R}^K$ is distributed as a multivariate normal
\begin{equation}
    \eta_i \sim \text{MvN} \left( \mu_p, \Sigma_p \right),
\end{equation}

with  hyperparameters $\mu_p \in  \mathbb{R}^{K-1}$ and  $\Sigma_p \in
\mathbb{R}^{K-1\times  K-1}$.  To   preserve  identifiability  on  the
simplex  we fix  $\eta_{iK} =  0  ~ \forall  i$, that  is, only  $K-1$
parameters  are  free. By  using  a  logit  multivariate normal  prior
instead  of a  typical Dirichlet  prior we  allow for  correlations to
exist  between individual  membership scores  \citep{gruhl14}, further
enhancing the flexibility of the BPM.

In what follows we describe the priors that are used for the variables
of the  model. We decompose  the covariance matrices of  the component
distributions $\Sigma_k ~  k \in 1,2,3$ and of  the multivariate prior
$\Sigma_p$ as

\begin{equation} \label{cov}
\Sigma = \begin{pmatrix} \sigma_{1} & 0 \\ 0 & \sigma_{2} \end{pmatrix} C \begin{pmatrix} \sigma_{1} & 0 \\ 0 & \sigma_{2} \end{pmatrix},
\end{equation}

where  $C$  is a  $K-1  \times  K-1$  symmetric and  positive-definite
correlation matrix. For the correlation matrix we use a LKJ prior with
one degree  of freedom, which corresponds to  the uniform distribution
for correlation  matrices. For  the standard deviations  a Half-Cauchy
prior  with a  scale  parameter $\gamma  =  5$ is  used.  The LKJ  and
Half-Cauchy are  standard choices  for weakly informative  priors. The
mean   vector    of   the   components   is   set    as   defined   in
Sect. \ref{spm_db}. Finally, for  $\mu_p$ a normal prior with standard
deviation $\sigma=10$ is  used. The mean hyperprior of  $\mu_p$ is set
to  $(2, 1)$  which translates  into to  $(0.66, 0.24,  0.09)$  in the
simplex, corresponding  to our  expectation that the  concentration of
the  bulge is  larger  than the  disk  and the  disk  larger than  the
cluster.

We  are  interested  in  obtaining distributions  for  the  membership
vectors, the components covariance,  and their respective priors given
the  data. In  traditional  Bayesian settings,  distributions for  the
parameters are  obtained via approximating the joint  posterior of the
model using Markov  chain Monte Carlo (MCMC) methods.  But MCMC scales
poorly  with  high  complexity  and  high  dimensionality  models,  as
multi-modality  and  non-identifiability  issues  may  slow  or  stall
convergence  of the  chain. Due  to this  we opt  for  an optimization
approach  based  on  automatic differentiation  variational  inference
ADVI. With ADVI one fits the posterior with a much simpler variational
distribution  by  minimizing  their Kullback-Liebler  (KL)  divergence
using     gradient-descent     based     routines.     Contrary     to
expectation-maximization, ADVI is fully  Bayesian in the sense that we
can draw  from the variational posterior and  obtain distributions for
the parameters rather than point estimates.

The  hierarchical   model  is  programmed  using   the  PyMC3  package
\citep{salvatier16}, which includes an ADVI routine. Because the model
is continuous  and differentiable  on its parameters  we can  use ADVI
straightly.  We consider  that the  procedure has  converged  when the
difference  in successive KL  divergences is  smaller than  0.5\%. The
results of this procedure are presented in Sect. \ref{membshipprob}.


\end {document}